# Numerical simulations of the mutual effect among the superconducting constituents in a levitation system with translational symmetry

Guang-Tong Ma[1,a)], Huan Liu[1,2], Xing-Tian Li[1,2], Han Zhang[1,2], and Yuan-Yuan Xu[1]

[1]*Applied Superconductivity Laboratory, State Key Laboratory of Traction Power, Southwest Jiaotong University, Chengdu, Sichuan 610031, China*

[2]*School of Electrical Engineering, Southwest Jiaotong University, Chengdu, Sichuan 610031, China*

## ABSTRACT

By the introduction of a generalized magnetic vector potential, which contains the contributions of both the magnetic and electric parts, and the use of the Ampere's law within the quasistatic approximation as the state equation, the partial differential equations for governing the electromagnetic properties of superconductors as well as the surrounding coolant were established and numerically discretized by resorting to the finite-element technique and finite-difference scheme, respectively, in the spatial and temporal domain. In conjunction with an analytic method to calculate the magnetic field generated by permanent magnet, we compiled a numerical tool for performing an intricate study of the mutual effect among the superconducting constituents in a superconducting levitation system with translational symmetry. Taking a superconducting unit with three constituents inside as a practice, we simulated the electromagnetic responses of this unit while moving in the nonuniform magnetic field generated by permanent magnet guideway and, identified the influences of the mutual effect on the levitation force as well as on the distributions of the magnetic flux density, the supercurrent density, and the levitation force density by comparing to an *envisaged* reference, one constituent was simulated with all the rest absent to remove the mutual effect. The insights attained by the present study, mostly being inaccessible from the experiments, are aimed to provide useful implications for the design of a superconducting levitation system for the transit and analogous purposes, which usually employ multiple superconductors to achieve the desired capability.

## I. INTRODUCTION

The self-stable levitation in physics realized by releasing a magnet over a superconductor or vice versa[1] is deemed promising elements for a wide range of applications such as the transit[2–5] and its analogue.[6,7] An important characteristic regarding these uses is the electromagnetic responses, particularly the magnetic forces, of the superconductor while moving in the nonuniform magnetic field generated by the magnet devices. For this purpose, from a theoretical point of view, in the early phase after the discovery of this self-stable levitation,[1] the electromagnetic responses of the superconductor were basically explored in the framework of the mirror-image model by considering the superconductor as a perfect diamagnet,[8–10] which is easy to be implemented but its applicable scope is limited due to the dipole approximation, though its extended form to reflect the hysteresis of magnetic forces has been proposed in recent years.[11]

To investigate the electromagnetic responses of the superconductor in a more realistic sense, various electromagnetic models derived from the Maxwell's equations have been established and verified in both two- and three-dimensional levels.[12–19] As far as our considerations about the levitation system with superconductors over a permanent magnet guideway (PMG) is concerned, using these models, the geometrical and material effects on the levitation performance,[12,20] and the distributions of the magnetic field and the induced supercurrent[13,17,21] have been studied, apart from the influence of the lateral movement on the levitation force.[22,23] Also, the propitious configuration of PMG for improving the levitation capability has been estimated[14,24,25] and miscellaneous strategies for the practical design of such levitation system have been suggested.[14,26–30]

However, none of these aforementioned efforts were made to the superconducting unit containing multiple constituents, a practically employed situation due to the limit of the dimension of a single superconductor currently attainable.[4] The induced magnetic field generated by one constituent in a superconducting unit will definitely redistribute the applied magnetic field in the domain occupied by the rest, causing the constituents to be magnetically coupled and thus mutually influenced. Consequently, the electromagnetic responses of a superconductor in a unit having multiple constituents are anticipated to be distinct as compared to those of it alone, entailing the study of the superconducting unit with multiple constituents. Since its importance, recently has this aspect been concerned towards a superconducting unit made up of two or three infinite constituents basing on the critical-state approximation and a magnetic-energy minimization procedure.[31,32] Although some insights have been observed by these investigations, taking the field-dependent feature of critical current density, which may be particularly critical for the magnetically coupled constituents, into account and identifying how the mutual effect due to this magnetic coupling among the superconducting constituents alters the electromagnetic responses in comparison with that without mutual effect remain open for better understanding the electromagnetic behavior of the levitation system with multiple superconductors over a PMG.

The superconducting unit in this work is presumed to be made up of three constituents with a small gap in between, all being translational invariant over an infinitely extended PMG and understood as the Y-Ba-Cu-O bulks cooled with liquid nitrogen. The mutual effect among the constituents is taken into account during the simulations by means of numerically discretizing these constituents and then calculating them as a whole. Meanwhile, an *envisaged* case of considering one constituent with all the rest absent is also estimated with the spatial mesh reserved, serving as a reference to the *actual* case with all constituents present, for identifying how the mutual effect among the constituents acts.

This paper is structured as follows. In Sec. II we describe the general framework of the mathematical foundations for governing the electromagnetic field of the levitation system with a single or multiple superconductors, which lays the basis of our numerical simulations. Making recourse to these, in Sec. III we first introduce two different configurations of PMG and visualize the calculated distributions of magnetic flux lines generated by these PMGs, and then, present and discuss the numerical results procured for the levitation force as well as for the spatial distributions of the magnetic flux density, the supercurrent density and the levitation force density, to unveil how the mutual effect affects these electromagnetic quantities. The conclusion is presented in Sec. IV. Mathematical formulas for analytically calculating the magnetic field of a permanent magnet (PM) can be found in the Appendix.

## II. MATHEMATICAL FOUNDATIONS

### A. Governing equations of electromagnetic field

We introduce a generalized magnetic vector potential and use the Ampere's law within the quasistatic approximation as the state equation to establish the partial differential equation for governing the electromagnetic field of the levitation system with a single or multiple superconductors. The prominent advantage of this course is that, for such a two-dimensional problem concerned in this work, only the vector potential along the direction of translational invariance (*x*-axis in a Cartesian coordinate shown in Fig. 1) needs to be defined and solved, which is rather profitable in terms of reducing the number of degrees of freedom when adopting the finite-element technique to discretize the spatial domain including the superconductors and the surrounding coolant as well. We postulate the induced field, generated by the supercurrent, does not influence the coercive field of the PM, which allows decoupling the calculation of the PMG and the superconductors.

Recalling the Faraday's law and the definition of $\mathbf{B} = \nabla \times \mathbf{A}$, the nontrivial component of the induced electric field $E_x$ for a single superconductor can be expressed as

$$E_x = -\frac{\partial (A_{sc,x} + A_{ex,x})}{\partial t} - (\nabla V)_x, \tag{1}$$

with the magnetic vector potential along the x-direction $A_x$ being decomposed into $A_{sc,x}$ and $A_{ex,x}$, where $A_{sc,x}$ represents the vector potential induced by the supercurrent, whereas $A_{ex,x}$ serves as the magnetic excitation generated by the PMG. The gradient of the electric scalar potential $\nabla V$ is invariant across the yz-plane and only depends on time t, as both **E** and **A** are nontrivial merely along the x-axis and independent of the x-axis.[33] If we denote the value of $(\nabla V)_x$, at an arbitrary time instant $t_n$, as $C(t_n)$, Eq. (1) can be rewriten as,

$$E_x = -\frac{\partial \left( A_{sc,x} + \int_0^{t_n} C(t) dt + A_{ex,x} \right)}{\partial t}, \tag{2}$$

Making use of the Ampere's law within the quasistatic approximation and exploiting Eq. (2), the electromagnetic properties in the superconductor can be governed by

$$-\frac{1}{\mu_0} \left( \frac{\partial^2 A_{sc,x}}{\partial y^2} + \frac{\partial^2 A_{sc,x}}{\partial z^2} \right) + \sigma \frac{\partial \left( A_{sc,x} + \int_0^{t_n} C(t) dt \right)}{\partial t} + \sigma \frac{\partial A_{ex,x}}{\partial t} = 0, \tag{3}$$

where we have assumed the magnetic field intensity and magnetic flux density in the superconductor is related linearly with the vacuum permeability $\mu_0$.

Since $\int_0^{t_n} C(t) dt$ is independent of y and z, it is expedient to adopt a generalized vector potential $A'_{sc}$, which defines as

$$A'_{sc,x} = A_{sc,x} + \int_0^{t_n} C(t) dt, \tag{4}$$

to make the integral term in Eq. (2) implicit. Eq. (3) is thus reduced to

$$-\frac{1}{\mu_0}\left(\frac{\partial^2 A'_{sc,x}}{\partial y^2}+\frac{\partial^2 A'_{sc,x}}{\partial z^2}\right)+\sigma\frac{\partial A'_{sc,x}}{\partial t}+\sigma\frac{\partial A_{ex,x}}{\partial t}=0 \tag{5}$$

where $A'_{sc,x}$ is the unknown to solve and the electric conductivity of superconductor is strongly dependent on the solution $A'_{sc,x}$ as well as $A_{ex,x}$, which necessitates the action of numerical iteration. It is worth noting that, the contribution of the electric scalar potential will *stand out* in the computational region far from the superconductor, where the vector potential $A_{sc,x}$ due to the supercurrent is negligible, i.e., $A'_{sc,x}\cong \int_0^{t_n} C(t)dt$ holds (if the superconductor is subjected to a symmetric magnetic field, $C(t_n)\equiv 0$ stands and it is not needed to introduce the generalized vector potential thereby).

For a superconducting unit containing $m$ constituents, the magnetic excitation to an arbitrary constituent $i$ is generated by the PMG as well as by all the rest, which requires the modification of the electromagnetic master equation for the constituent $i$ to be,

$$-\frac{1}{\mu_0}\left(\frac{\partial^2 A'_{sc,x,i}}{\partial y^2}+\frac{\partial^2 A'_{sc,x,i}}{\partial z^2}\right)+\sigma\left(\frac{\partial}{\partial t}\sum_{j=1}^{m}A'_{sc,x,j}-\frac{\partial}{\partial t}\sum_{j=1,j\neq i}^{m}A'_{sc,x,j,\Gamma}\right)+\sigma\frac{\partial A_{ex,x}}{\partial t}=0 \tag{6}$$

where $A'_{sc,x,j,\Gamma}$ refers to the boundary unknown of the constituent $j$ ($j\neq i$), where $A'_{sc,x,j,\Gamma}\cong (\nabla V)_{x,j}$ stands. In this case, $m$ different unknowns, $A'_{sc,x,1}$, $A'_{sc,x,2}$, ..., $A'_{sc,x,m}$ should be defined at each finite element node, and they are coupled through the second term of the left hand side of Eq. (6).

The coolant to provide a cryogenic environment for the superconductors is supposed to be dielectric and the second term of the left hand side of Eqs. (5) and (6) thus vanishes in the coolant. The electromagnetic master equation in the coolant is actually represented by the Laplace's equation. The continuity of $A'_{sc,x}$ as well as its normal derivative is applied to the interior bounds between the superconductor and the coolant and, the value of $A'_{sc,x}$ on the outer bounds is *not constrained*.

The nonlinear feature of *J–E* relation in the superconductor is characterized by a smoothed Bean–Kim's model of the critical state in the hyperbolic tangent approximation[18,34]

$$J_x=J_{c0}\left(\frac{B_0}{|\mathbf{B}|+B_0}\right)\tanh\left(\frac{E_x}{E_0}\right)=J_{c0}\left(\frac{B_0}{|\nabla\times\mathbf{A}|+B_0}\right)\tanh\left(-\frac{\partial\left(A'_{sc,x}+A_{ex,x}\right)}{\partial t}\frac{1}{E_0}\right), \tag{7}$$

where $J_{c0}$ is the critical current density in the absence of magnetic field, $E_0$ is the characteristic electric field and $B_0$ represents the critical magnetic flux density for which $J_c = J_{c0}/2$.

## B. Numerical implementation

Supported by previous experience,[18,34] the discretization of spatial domain using the finite-element technique is executed taking Galerkin's method[35] into account, whereas the discretization of temporal domain is performed on the basis of the finite-difference technique via the backward Euler's scheme.[36] The final nonlinear system of finite-element equation to resolve takes the form:

$$\left( [\mathbf{K}(\mu_0)] + \frac{[\mathbf{Q}^n(\sigma)]}{\Delta t} \right) \{A_{sc,x}^{\prime n}\} = \frac{[\mathbf{Q}^n(\sigma)]}{\Delta t} \{\{A_{ex,x}^{n-1}\} - \{A_{ex,x}^n\} + \{A_{sc,x}^{\prime n-1}\}\}, \tag{8}$$

where $\Delta t$ is the time interval between the successive time instants, and the superscript $n$ and $n$-1 represent, respectively, the vector/matrix for the current and last time instant.

A linear triangular nodal element is deployed to generate the entries of the stiffness matrix $[\mathbf{K}(\mu_0)]$ and the damping matrix $[\mathbf{Q}^n(\sigma)]$. The vector $\{A_{ex,x}\}$ is known at each time instant and serves as the stimulated term from the PMG, which is calculated by the analytic method described in the Appendix.

The Jacobian-free Newton-Krylov approach, founded on a synergistic combination of Newton-type methods for superlinearly convergent solutions of nonlinear equations and Krylov subspace methods for solving the Newton correction equations,[37] is called upon to treat the nonlinearity of Eq. (8) with the *associated algebraic equations after linearization solved by means of the generalized minimal residual algorithm*.[38] An invaluable advantage lie in the fact that such a course avoids the usual evaluation of the Jacobian matrix for each element, and saves massive demands on computer memory and processing time thereby.

According to Lorentz's equation, the magnetic force per length exerting on the superconductors along the $z$-direction $F_z$, i.e., the levitation force, and along the $y$-direction $F_y$, i.e., the guidance force, can be numerically calculated by respectively

$$F_z = \sum_{i=1}^{M} \iint_S J_x B_y \, dy dz = \sum_{i=1}^{M} \sum_{j=1}^{N} J_{x,j,i}^e B_{y,j,i}^e \Delta S_{j,i}^e, \tag{9}$$

$$F_y = -\sum_{i=1}^{M} \iint_S J_x B_z \, dy dz = -\sum_{i=1}^{M} \sum_{j=1}^{N} J_{x,j,i}^e B_{z,j,i}^e \Delta S_{j,i}^e, \tag{10}$$

where the superscript *e* denotes the value of parameters at each element meshed by finite-element technique, and $\Delta S$ represents the area of each element, and *M* is the amount of the superconductor and *N* is the number of mesh element in each superconductor. The magnetic flux density, $B_y$ and $B_z$, includes the contribution of the PMG as well as of all the superconductors. Worthy of mention here is that, the self-field induced in each superconductor, though has actually no contribution to its own levitation and guidance forces as a whole, it can certainly affect the spatial distribution of its own forces, and must be therefore included in $B_y$ and $B_z$ to obtain the accurate distributions of the levitation force presented in Fig. 7 in Sec. III.

## III. RESULTS AND DISCUSSION

We considered two different configurations of PMG, both of which are derived from the Halbach array.[39] The contour lines of magnetic flux of the horizontal component $B_y$, calculated by the analytic model described in the Appendix, is delineated for each PMG and displayed in Fig. 1, alongside the related geometrical and material parameters. The PMGs, both containing five PMs with varied directions of magnetization following the Halbach array, have the ability to concentrate the magnetic field mostly above the upper space as depicted in Fig. 1, resulting in the symmetry of magnetic field only occurs in terms of the vertical direction and most importantly, an improved levitation performance.[40]

Only the calculated results of the zero-field-cooling (ZFC) condition, with the magnetic field absent during the transition to the superconductive state, is presented in this work. The ZFC condition is imitated in such a way that the initial entries of stimulated term in Eq. (8) are null, i.e., $\{A_{ex,x}^0\} = 0$, with a high starting position relative to the PMG to represent the physical fact thereof and the initial condition of unknown $\{A_{SC,x}^0\}$ is specified to be null, i.e., $\{A_{SC,x}^0\} = 0$.

In the following simulations, we assign the dimension of each constituent in the superconducting unit is 30 mm in width and 15 mm in thickness, and the gap between the constituents is 2 mm. We have used $J_{c0} = 1.5 \times 10^8$ A/m²,[4] $E_0 = 5 \times 10^{-6}$ V/m,[18] and $B_0 = 0.25$ T,[41] to roughly represent the material properties of bulk Y-Ba-Cu-O at 77 K. According to the geometrical and material characteristics of the PMGs proposed above, the starting position of vertical movement is 120 mm to approach the ZFC condition over both PMGs, where the strength of applied magnetic field is negligible. The minimum vertical distance between the superconductors and the PMG is 5 mm and the speed of the superconductors is 1 mm/s to create a quasistatic state, for all cases estimated below.

### A. Mutual effect on the levitation force

Figure 2 portrays the hysteretic loops of the levitation force on each individual constituent for the

7 / 22

*actual* case as well as for the *envisaged* case. The levitation force of the right constituent in both cases is invisible due to the reason of its symmetry to the left one. The development of the guidance force as the vertical movement, on each individual constituent of the *actual* case, was also plotted as an inset in each figure.

The main finding upon this figure include: (i) the levitation force on the left constituent over PMG_A and on the middle constituent over PMG_B, whose domains are dominated by the vertical component of magnetic field, is reduced due to the mutual effect, whereas the levitation force on the middle constituent over PMG_A and on the left constituent over PMG_B, whose domains are dominated by the horizontal component of magnetic field, is enhanced; (ii) as a consequence of the mutual effect, the levitation force on both constituents over both PMGs tends to be a finite value but not zero during the movement to its initial position, which is obviously different from that of the *envisaged* case. This residual levitation force is generated by the magnetic interaction of the induced supercurrent in one constituent with the induced magnetic field of the rest; (iii) regardless of the configuration of the PMG, the guidance force of the middle constituent is null but the left and right constituents undergo a finite guidance force with opposite direction. The superconducting unit as a whole can achieve a null guidance force.

Though the mutual effect brings out a pronounced alteration of the levitation force on each individual constituent, the levitation force as a whole, whose hysteretic loops of over both PMGs were plotted in Fig. 4, is influenced negligibly due to the mutual effect, only a slight increase of the hysteretic loop over PMG_B being found. The levitation force for the *envisaged* case in Fig. 3 was obtained by summating the contributions from each superconductor calculated individually.

## B. Mutual effect on the distribution of magnetic flux density

The distributions of magnetic flux density *B* inside the superconductors, at the time instant when the smallest vertical distance was achieved during the calculations of the hysteretic loops of levitation force in Fig. 2, was visualized and presented in Fig. 4 for both PMGs in the ZFC condition, being the intensity of the red color proportional to the strength of the local flux density.

Generally, the distributions reveal a physical symmetry of magnetic flux density in terms of the perpendicular bisector of the middle constituent in each case and, there exists a region, for all constituents of both *actual* and *envisaged* cases, where no magnet flux penetrates, i.e., the flux-free region.

Specifically speaking, (i) for the case with superconductors over PMG_A shown in Fig. 4 (a), the mutual effect on the middle constituent brings out a notable decrease of the crescent-shaped flux-free region and extends the strong flux-penetrated region at the bottom slightly upwards. Due to the mutual effect, the strong flux-penetrated region localized at the lower-inner-corners of the side constituents is slightly suppressed and the flux-free region of the side constituents, formed in the *envisaged* case, is modulated to be quasi onion-shaped one, by mostly altering the flux-penetrated region at the inner parts. The flux lines, mostly in the adjacent region between the constituents, are distorted by the mutual effect; (ii) for the case with superconductors over PMG_B shown in Fig. 4 (b), as a consequence of the mutual effect, the onion-shaped flux-free region formed in the middle constituent of the *envisaged* case is markedly reduced and the strength of the flux-penetrated region localized at the lower-corners is weakened. The mutual effect on the side constituents gives rise to a broader region of flux penetration, or in other words, a smaller flux-free region, and slightly extends the strong flux-penetrated region at the bottom upwards. The flux lines, mostly in the adjacent region between the constituents, are also distorted by the mutual effect, similarly to that found in the case of over PMG_A.

## C. Mutual effect on the distribution of supercurrent density

The distributions of supercurrent density $J_x$ inside the superconductors, at the time instant when the smallest vertical distance was achieved during the calculations of the hysteretic loops of levitation force in Fig. 2, was visualized and presented in Fig. 5 for both PMGs in the ZFC condition, being the intensity of the red and blue color proportional to the strength of the plus and minus supercurrent respectively.

Generally, the distributions reveal a physical symmetry of supercurrent density in terms of the perpendicular bisector of the middle constituent over PMG_A and a physical antisymmetry for the case over PMG_B. In each picture, there exists a region, where no supercurrent flows to be in concert with the flux-free region in Fig. 4, and the whole domain of each constituent in the *actual* and *envisaged* cases, is naturally divided into two regions, one flowing supercurrent opposite to the other. The two bands with the highest density in each constituent, one flowing the supercurrent opposite to the other, are always formed near the flux-free region, where the variation of the magnet flux is intensive according to Fig. 3.

Specifically speaking, (i) for the case with superconductors over PMG_A shown in Fig. 5 (a), the mutual effect shortens the band with the highest density in the middle constituent, and results in a reduced current-free region. Due to the mutual effect, the current-free regions formed in the side constituents of the *envisaged* case tends to be onion-shaped by mostly redistributing the plus supercurrent in the inner part of each side constituent; (ii) for the case with superconductors over PMG_B shown in Fig. 5 (b), the mutual effect on the middle constituent gives rise to a reduced current-free region, and extends the bands with the highest density inward. As a result of the mutual effect, the current-free region of the side constituents is slightly suppressed, as that found in the middle constituent of over PMG_A.

The evolution of the net current for the three constituents in both *actual* and *envisaged* cases as a function of the vertical distance over both PMGs was sketched and presented in the Fig. 6. This figure indicates that, the induced supercurrent of each constituent in the *actual* case, estimated by integrating the supercurrent density across the respective cross section, is naturally zero, or very close to zero due to the numerical error, over both PMGs, as that found in the *envisaged* case, demonstrating that the electrically isolated connection among the constituents is virtually left unchanged with the mutual effect present. This finding also proves that, with the mathematical foundations derived above, the zero net current can be automatically fulfilled without the constraint of supercurrent used in Ref. 32.

### D. Mutual effect on the distribution of levitation force density

The distributions of levitation force density inside the superconductors, at the time instant when the smallest vertical distance was achieved during the calculations of the hysteretic loops of levitation force in Fig. 2, was visualized and presented in Fig. 7 for both PMGs in the ZFC condition, being the intensity of the red and blue color proportional to the strength of the plus and minus force density respectively. In the following description, each superconductor in Fig. 7 is divided into two domains, one is the repulsive domain with a plus density, positively contributing to the levitation force, and the other is the attractive domain with a minus density, negatively contributing to the levitation force.

Generally, the distributions reveal a physical symmetry of the levitation force in terms of the perpendicular bisector of the middle constituent in each case. Regardless of the PMG, there exists a force-free region, where the contribution to the levitation force is null, for both the *actual* and *envisaged* cases, and the attractive domain is usually formed at the upper part of the superconductor. The highest density of both the repulsive and attractive domains is always achieved in the superconductors subjected to a magnetic field with horizontal component dominated.

Specifically speaking, (i) for the case with superconductors over PMG_A shown in Fig. 7 (a), the mutual effect on the middle constituent gives rise to a reduced force-free region and, owning to the mutual effect, the strength of the attractive and repulsive domain is respectively degraded and upgraded. Due to the mutual effect, the force-free region in the side constituents is reshaped and the contour lines of force, mostly in the adjacent region between the constituents, are distorted by the mutual effect; (ii) for the case with superconductors over PMG_B shown in Fig. 7 (b), the force-free region in the middle constituent is slightly shrunk as a result of the mutual effect and the attractive domain is gently broadened. Similarly to those found over PMG_A, the force-free region in the side constituents is reduced due to the upward expansion of the repulsive domain and the inward extension of the attractive domain. The contour lines of force, mostly in the adjacent region between the constituents, are also distorted by the mutual effect.

## IV. CONCLUSIONS

Considering a superconducting levitation system with translational symmetry, numerical simulations of the mutual effect among the superconducting constituents therein have been carried out by introducing a generalized magnetic vector potential within the quasistatic approximation. The investigations of the mutual effect on the levitation force as well as on the distributions of the magnetic flux density, the supercurrent density and the levitation force density, with the superconductors over two promising PMGs, derived from the Halbach array, lead to the main findings as follows.

Providing the geometrical and material characteristics of both PMG and superconductors chosen in this work, the mutual effect brings out a significant enhancement of the levitation force on the constituents, whose domains are dominated by the horizontal component of magnetic field, and conversely, the levitation force on the constituents, whose domains are dominated by the vertical component of magnetic field, is reduced due to the mutual effect. The influence of the mutual effect on the levitation force as a whole is however negligible over both PMGs. As a consequence of the mutual effect, the flux-free or current-free region inside the superconductors over both PMGs is essentially changed, accompanied by a transformation on the band with the highest density of supercurrent, and the contour line of magnetic flux, particularly in the adjacent region between the constituents, is distorted. These intricate alterations cause a redistribution of the repulsive and attractive domain of levitation force in the superconducting constituents over both PMGs.

## ACKNOWLEDGEMENTS

The authors would like to thank the reviewers whose suggestions and comments are crucially valuable for improving this manuscript. This work was supported in part by the National Natural Science Foundation of China under Grant 51007076, by the Fundamental Research Funds for the Central Universities under Grant SWJTU11ZT34, and by the State Key Laboratory of Traction Power at Southwest Jiaotong University under Grant 2013TPL_T05.

## APPENDIX: CALCULATION OF PM FIELD

We calculate the nonuniform magnetic field, generated by the PMG through the combined contribution of each PM therein, via an analytic model established by resorting to the surface current model.[14,42,43] Analyzing the calculation model for all PMs of a given PMG can be rather cumbersome, and we therefore focus on a vertically magnetized PM as a representative. The cross-sectional view of the surface current model for the designated PM of width $2w$ and thickness $d$ is schematically drawn in Fig. 8, with two sheet currents counter flowing near the marginal parts. If the PM is supposed to be magnetized uniformly with magnetization $\mathbf{M} = M_0 \mathbf{z}$, the volume current density is null due to the zero-gradient of **M** across the *yz* plane and only the surface current density, estimated by $\mathbf{j}_s = \mathbf{M} \times \mathbf{n}$, is left. The surface current density is respectively equal to $M_0\mathbf{x}$ and $-M_0\mathbf{x}$ for the left and right parts, as marked in Fig. 8. In this way, the problem is reduced to solve the magnetic field of two infinitely long current sheets of height *d* with opposite direction and separated by a dimension of 2*w*. In this case, the magnetic vector potential **A**, defined as $\mathbf{B} = \nabla \times \mathbf{A}$, at point (*y*, *z*) due to the combined contribution of the sheet current element $dI_1 = j_s dz'$ of the left one and $dI_2 = -j_s dz'$ of the right one is given by

$$d\mathbf{A} = \left( \frac{\mu_0 M_0}{4\pi} \ln \frac{(y-w)^2 + (z-z')^2}{(y+w)^2 + (z-z')^2} dz' \right) \mathbf{x}. \tag{A1}$$

Through the integral operation upon (A1) from $z' = 0$ and $z' = -d$, we formally arrive at the analytic equation to calculate the magnetic vector potential $A_x$ at point (*y*, *z*) generated by the PM in Fig. 8,

$$A_x = \frac{\mu_0 M_0}{4\pi} \left[ (z-z') \ln \frac{(y+w)^2 + (z-z')^2}{(y-w)^2 + (z-z')^2} + 2(y+w) \arctan \frac{z-z'}{y+w} - 2(y-w) \arctan \frac{z-z'}{y-w} \right]_{-d}^{0}. \tag{A2}$$

Once the expression of the magnetic vector potential is obtained, the two components of the magnetic flux density, $B_y$ and $B_z$, can be deduced to be

$$B_y = \frac{\mu_0 M_0}{4\pi} \ln \frac{\left[ (y+w)^2 + z^2 \right]\left[ (y-w)^2 + (z+d)^2 \right]}{\left[ (y+w)^2 + (z+d)^2 \right]\left[ (y-w)^2 + z^2 \right]},$$

$$B_z = \frac{\mu_0 M_0}{2\pi} \left( \arctan \frac{z}{y-w} + \arctan \frac{z+d}{y+w} - \arctan \frac{z}{y+w} - \arctan \frac{z+d}{y-w} \right), \tag{A3}$$

according to the relations that $B_y = \partial A_x / \partial z$, and $B_z = -\partial A_x / \partial y$.

The contribution of the PMs with other magnetization directions and locations in a certain configuration of PMG, such as those devised in Fig. 1, can be simply estimated by the geometrical operations of translation and/or rotation.

**Figures**

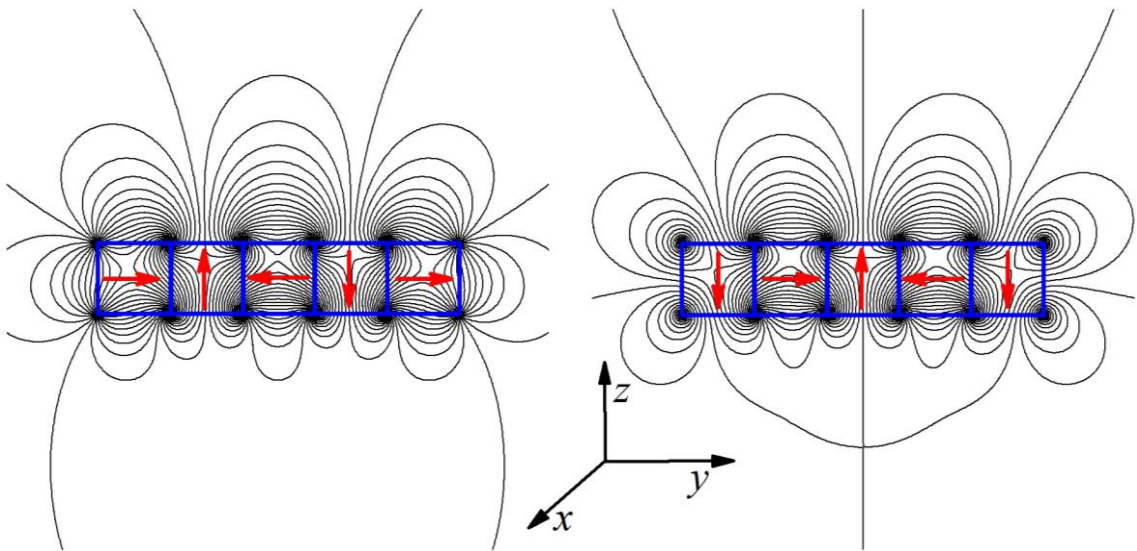

FIG. 1. (Color online) Contour lines of magnetic flux generated by the two different PMGs (hereafter marked as PMG_A and PMG_B from left to right), all with a geometrical configuration being invariant along the *x*-direction and symmetric in terms of the *z*-direction of a Cartesian coordinate system *x*, *y*, *z*. Shown for both plots is the horizontal component of magnetic flux density $B_y$. Each PM element, represented by rectangle with blue color in both PMGs, has an identical width and thickness being of 32 mm. A magnetization $M_0$ of being $8.753 \times 10^5$ A/m was assigned to all PM elements, intending to approximately reproduce the performance of an N35 magnet.

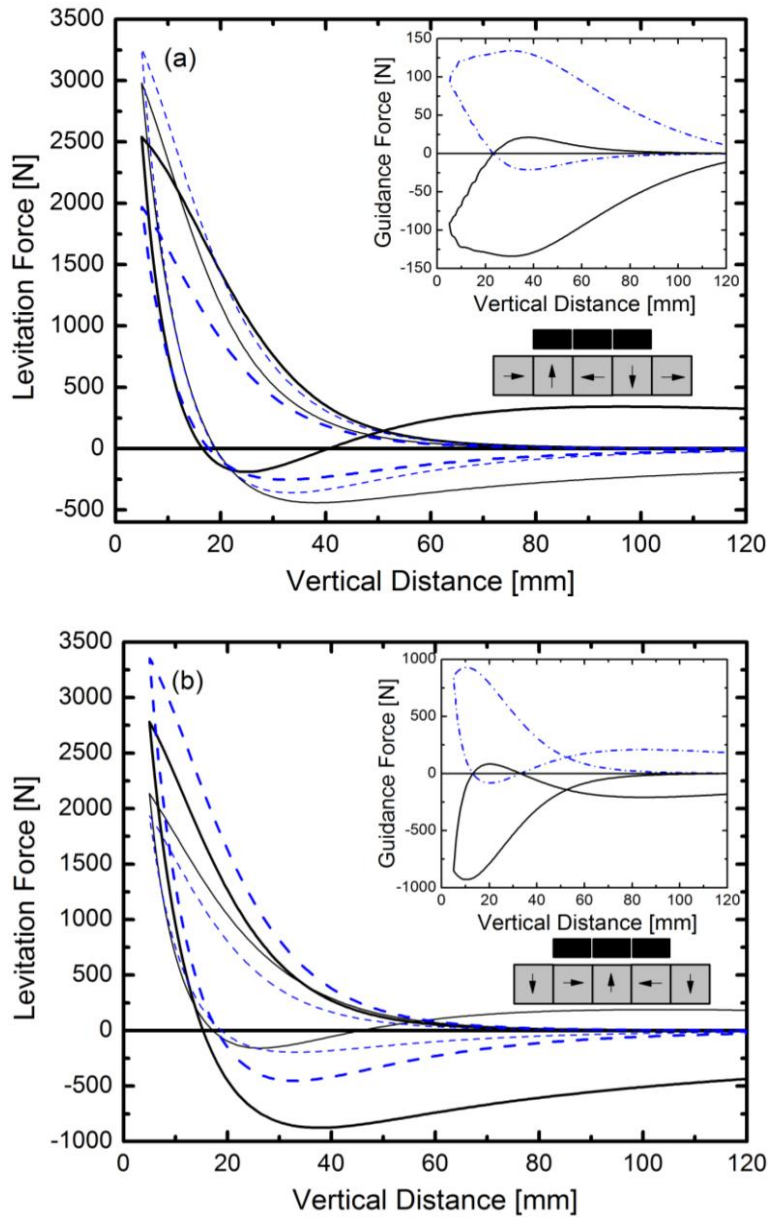

FIG. 2. (Color online) Hysteretic loops of levitation force on the left (thin solid line) and middle (thick solid line) constituent of a superconducting unit with three constituents inside while moving in the nonuniform magnetic field generated by respectively PMG_A (a) and PMG_B (b). Also shown for comparison by the corresponding dashed line is that obtained in the *envisaged* case. The concurrent hysteretic loops of guidance force on each constituent (left: solid line; middle: dash line; right: dash-dot line) of the *actual* case is plotted as an inset.

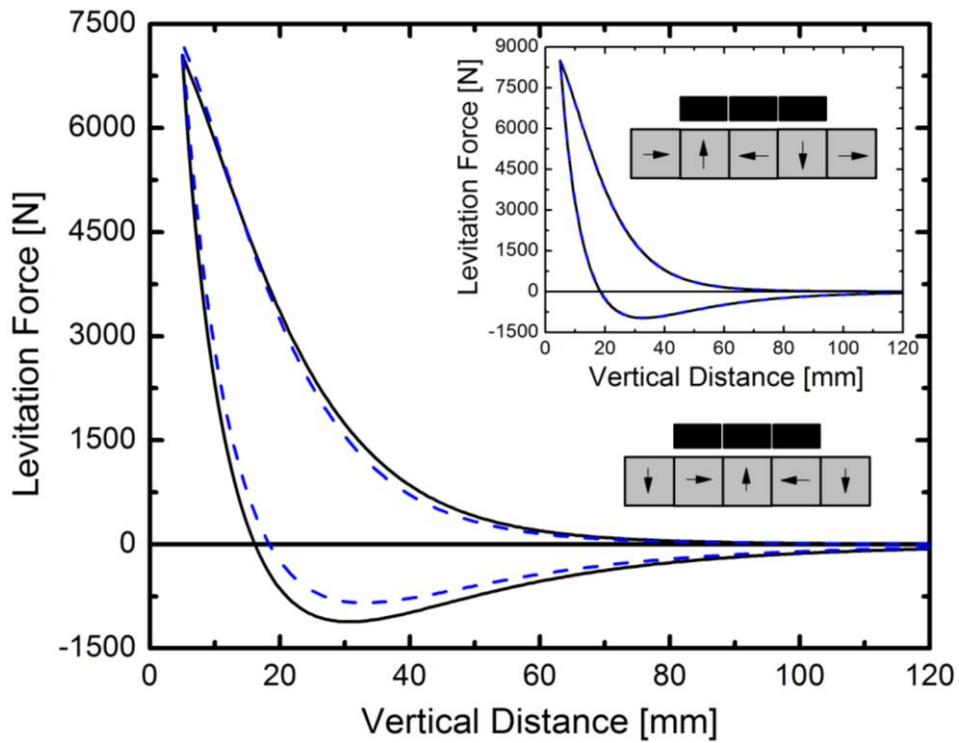

FIG. 3. (Color online) Hysteretic loops of the total levitation force on a superconducting unit with three constituents inside while moving in the nonuniform magnetic field generated by respectively PMG_A (inset) and PMG_B. Also shown for comparison is the summated result of levitation force on the individual superconductor obtained in the *envisaged* case (dash line).

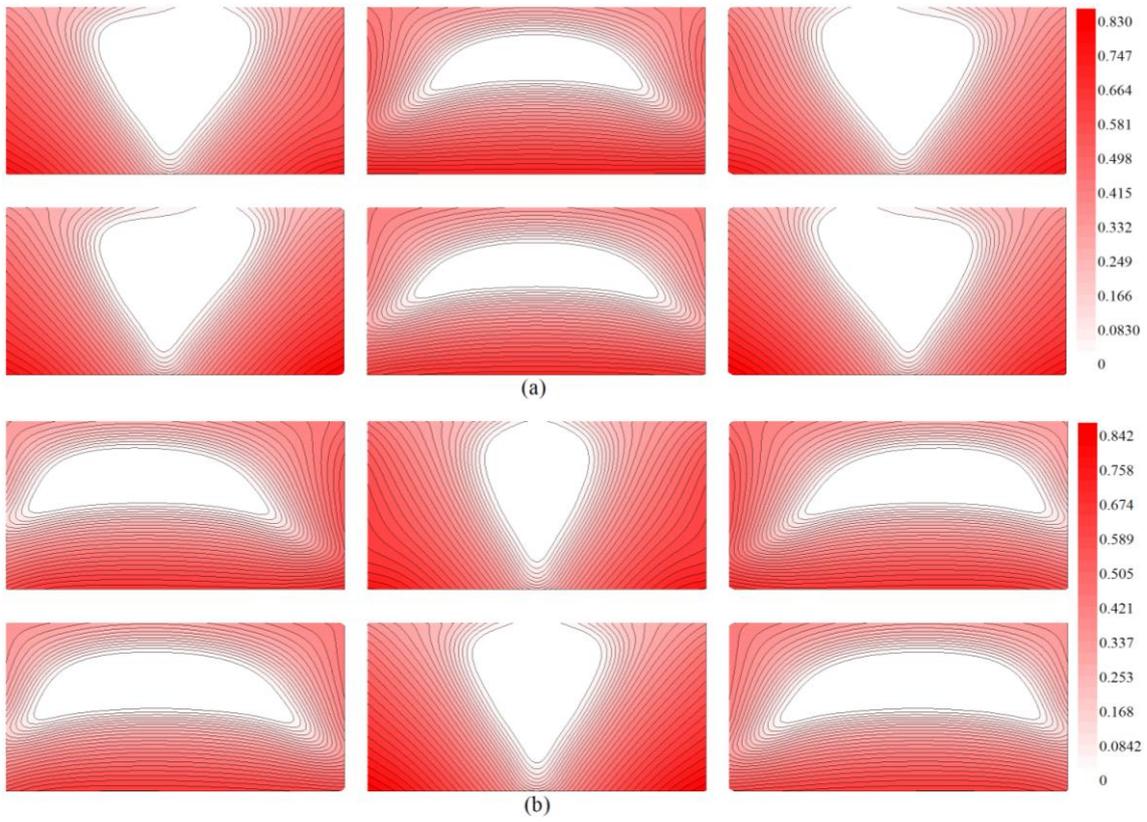

FIG. 4. (Color online) Snapshots of the distributions of magnetic flux density $B$ (T) inside the superconducting constituents at the time instant when the smallest vertical distance (5 mm) was achieved during the calculations of the hysteretic loops of levitation force in Fig. 2 over PMG_A (a) and PMG_B (b), referring to the proposed PMGs shown in Fig. 1. The lower part of each picture, shown for comparison, represents the distributions inside each individual superconductor in the *envisaged* case.

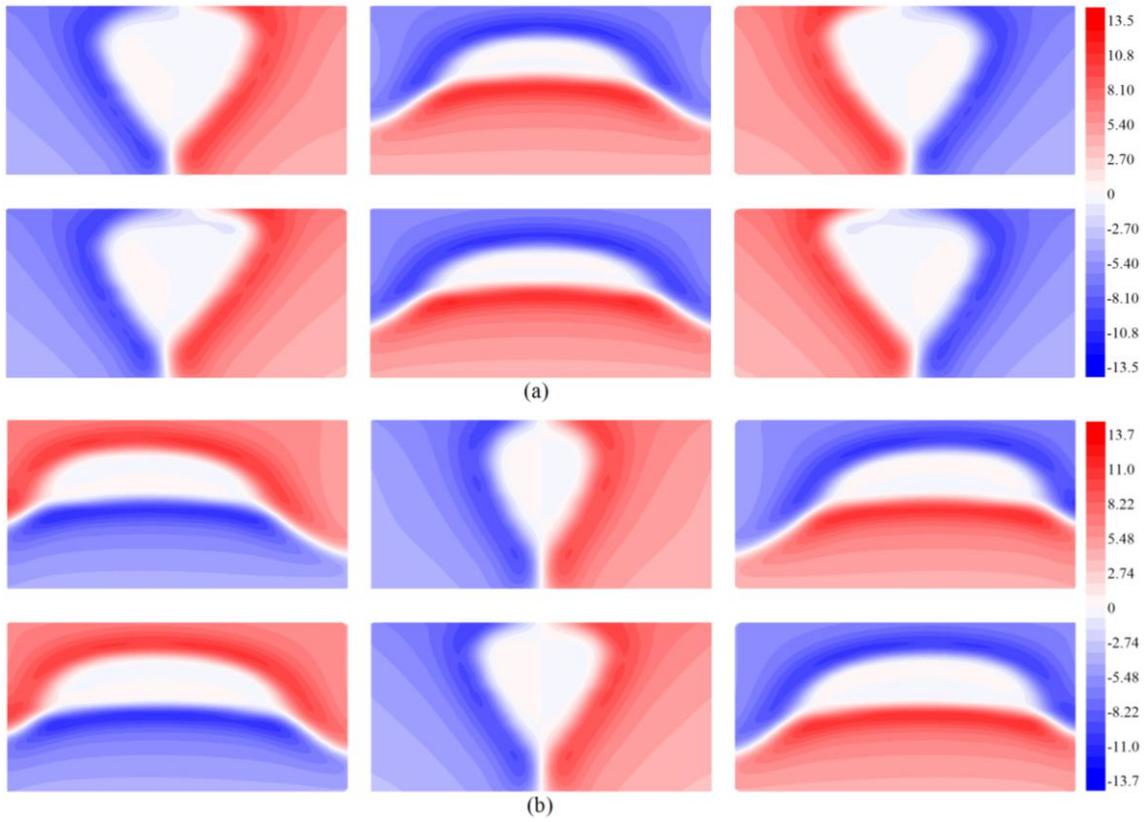

FIG. 5. (Color online) Snapshots of the distributions of supercurrent density $J_x$ ($10^7$ A/m$^2$) inside the superconducting constituents at the time instant when the smallest vertical distance (5 mm) was achieved during the calculations of the hysteretic loops of levitation force in Fig. 2 over PMG_A (a) and PMG_B (b), referring to the proposed PMGs shown in Fig. 1. The lower part of each picture, shown for comparison, represents the distributions inside each individual superconductor in the *envisaged* case.

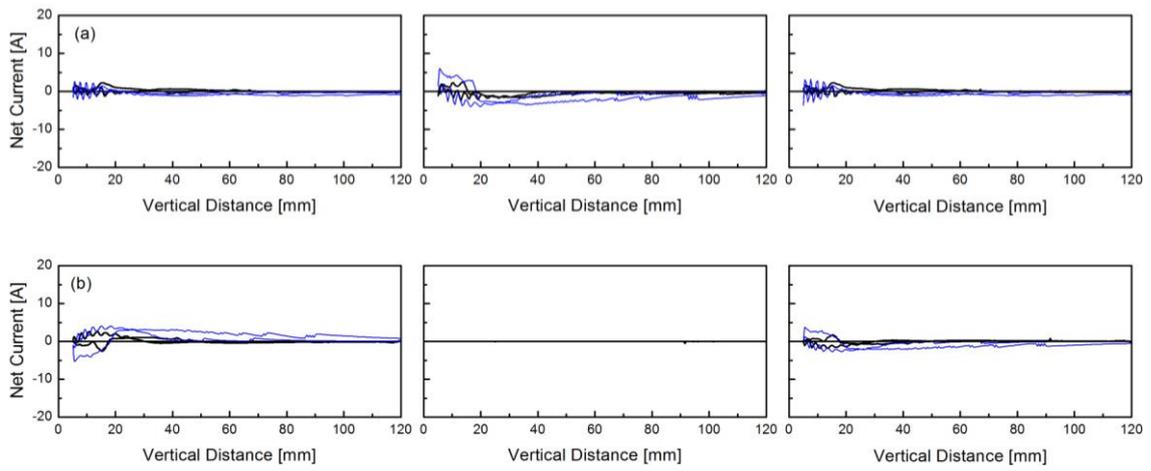

FIG. 6. (Color online) The evolution of the net current inside the three constituents as a function of the vertical distance for both *actual* (thick solid line) and *envisaged* cases (thin solid line) over PMG_A (a) and PMG_B (b).

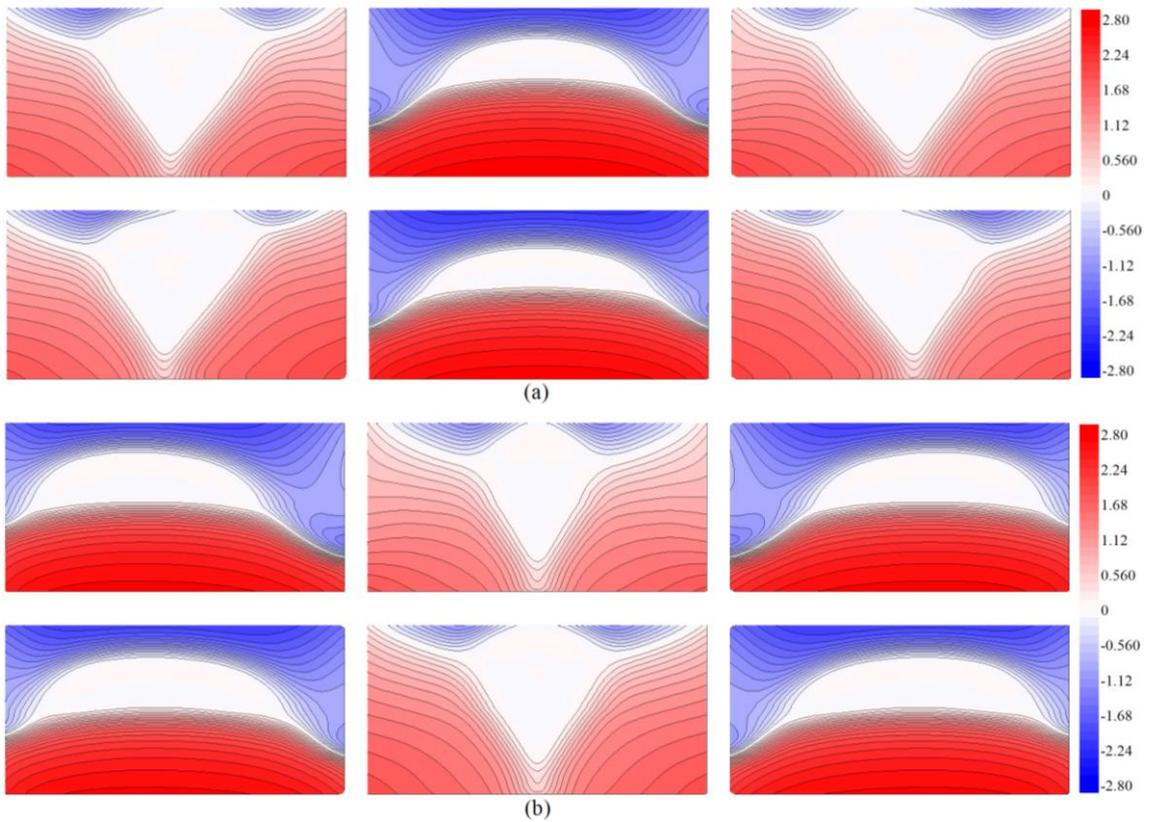

FIG. 7. (Color online) Snapshots of the distributions of levitation force density ($10^7$ N/m$^2$) inside the superconducting constituents at the time instant when the smallest vertical distance (5 mm) was achieved during the calculations of the hysteretic loops of levitation force in Fig. 2 over PMG_A (a) and PMG_B (b), referring to the proposed PMGs shown in Fig. 1. The lower part of each picture, shown for comparison, represents the distributions inside each individual superconductor in the *envisaged* case.

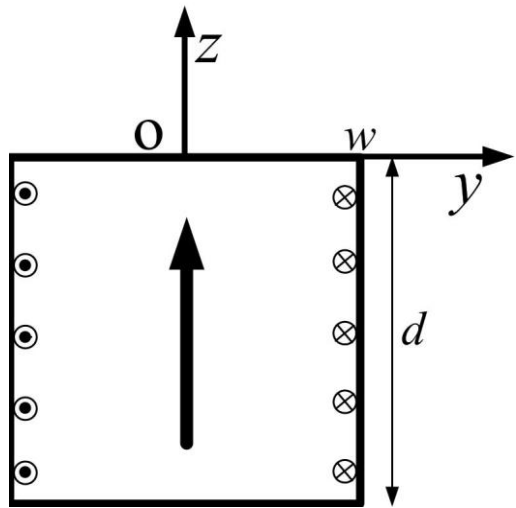

FIG. 8. Cross-sectional view of the surface current model for a vertically magnetized PM of width 2*w* and thickness *d* and being infinite along the invisible *x*-direction of a Cartesian coordinate system x, *y*, *z*.